\documentclass[aps,twocolumn,graphics,floatfix,tightenlines]{revtex4-2}
\usepackage{graphics}
\usepackage{epstopdf}
\usepackage{epsfig}
\usepackage{graphicx}
\usepackage{epsf,epic}
\usepackage{color}
\usepackage{subfig}
\usepackage{amsmath}
\usepackage{booktabs}
\usepackage{multirow}
\usepackage{physics}
\usepackage{hyperref}
\usepackage{amsfonts}
\usepackage{wrapfig}
\usepackage{breqn}
\usepackage{pstricks}
\usepackage[utf8x]{inputenc}
\usepackage{multirow}
\usepackage{fancyref}
\usepackage{pst-node}
\usepackage{float}
\usepackage{bm}
\usepackage{dcolumn}
\newcommand{\etal}{\textit{et al.\ }}
\newcommand{\ie}{\textit{i.e.\ }}

\makeatletter
\usepackage{etoolbox} 
\appto{\appendix}{%
	\@ifstar{\def\theequation@prefix{A.}}%
	{}%
}
\preto\maketitle{%
  \begingroup\lccode`~=`,
  \lowercase{\endgroup
  \let\saved@breqn@active@comma~
  \let~}\active@comma 
}
\appto\maketitle{%
  \begingroup\lccode`~=`,
  \lowercase{\endgroup
  \let~}\saved@breqn@active@comma 
}
\makeatother
\begin{document}
\title{Electronic structure and magnetism in $P4/nmm$ KCoO$_2$}
\author{Ozan Dernek,$^1$, Santosh Kumar Radha,$^{2}$ Jerome Jackson,$^3$ and Walter R. L. Lambrecht$^1$}
\affiliation{$^1$ Department of Physics, Case Western Reserve University, 10900 Euclid Avenue, Cleveland, OH-44106-7079}
\affiliation{$^2$ Agnostiq Inc., 325 Front St W, Toronto, ON M5V 2Y1}
\affiliation{$^3$ Scientific Computing Department, STFC Daresbury Laboratory, Warrington WA4 4AD, United Kingdom}
\begin{abstract}
  KCoO$_2$ has been found in 1975 to exist in a  unique structure with $P4/nmm$ spacegroup
  with Co in a square pyramidal coordination with the Co atoms in the plane linked by O in
  a square arrangement reminiscent of the cuprates but its electronic structure has not
  been studied until now. Unlike Co atoms in LiCoO$_2$ and NaCoO$_2$ in octahedral
  coordination, which are non-magnetic band structure insulators, the unusual coordination
  of $d^6$ Co$^{3+}$ in KCoO$_2$ is here shown to lead to a magnetic stabilization of an
  insulating structure with high magnetic moments of $4\mu_B$ per Co. The electronic band
  structure is calculated using the quasiparticle self-consistent (QS)$GW$ method and the
  basic formation of magnetic moments is explained in terms of the orbital decomposition
  of the bands. The optical dielectric function is calculated using the Bethe-Salpeter
  equation including only transitions between equal spin bands. The magnetic moments are
  shown to prefer an antiferromagnetic ordering along the [110] direction. Exchange
  interactions are calculated from the transverse spin susceptibility and a rigid spin
  approximation. The N\'eel temperature is estimated using the mean-field and Tyablikov
  methods and found to be between $\sim$100 and $\sim$250 K. The band structure in the AFM
  ordering can be related to the FM ordering by band folding effects. The optical spectra
  are similar in both structures and show evidence of excitonic features below the
  quasiparticle gap of $\sim$4 eV.
\end{abstract}
\maketitle
\section{Introduction}

  Among the alkali oxocobaltates, LiCoO$_2$ and NaCoO$_2$ have received much more
  attention than the larger cation ones because of their role in Li-ion batteries and the
  reported superconductivity in hydrated Na$_{1/3}$CoO$_2$:H$_2$O$_y$. Both of these
  exhibit the layered $R\bar{3}m$ structure, which can be viewed as consisting of edge
  sharing, octahedrally coordinated CoO$_2$ layers with a triangular Co-lattice stacked in
  an ABC stacking with intercalated Li or Na. The octahedral coordination, splitting
  $d$-levels in a six-fold degenerate $t_{2g}$ and a four-fold degenerate $e_g$ level,
  leads to a simple non-magnetic insulating band structure for the $d^6$ configuration of
  Co$^{3+}$ resulting from Li or Na donating their electron to the CoO$_2$ planes.
  However, starting with K, the alkali ions are too large to fit in this structure. Only
  half the amount of K can be maintained in between CoO$_2$ layer in this K$_x$CoO$_2$
  structure. For larger $x$, this structure becomes unstable and other structures were
  reported. In 1975, two different synthesis methods for KCoO$_2$ were reported and led to
  two totally different crystal structures with different Co-coordination. The first is a
  unique layered structure with square pyramidal coordination, with the $P4/nmm$ space
  group \cite{Jansen75}. The other is a stuffed cristobalite type structure in which Co is
  tetrahedrally coordinated with O in an open network of corner sharing tetrahedra, filled
  with K ions. Two related forms with space groups $I\bar{4}$ and $I\bar{4}2d$ were found
  and called respectively $\beta$ and $\alpha$-KCoO$_2$ \cite{Delmas75}. Besides these two
  papers, there seems to be no other studies reported on KCoO$_2$. Only very recently, a
  new synthesis method was developed for KCoO$_2$ in the pyramidal coordination and
  $P4/nmm$ spacegroup \cite{Frei2021}.

  Because of the occurrence of a square coordination of CoO$_2$, which resembles that of
  CuO$_2$ in high-$T_c$ materials, this phase may be of interest for non-conventional
  superconductivity. The occurrence of Co$^{3+}$ with a $d^6$ configuration in a pyramidal
  environment is also expected to show a high-spin, thus may have interesting magnetic
  properties. Here we present first-principles calculations of this material and its
  magnetic properties.

    \begin{figure}
      \includegraphics[width=4cm]{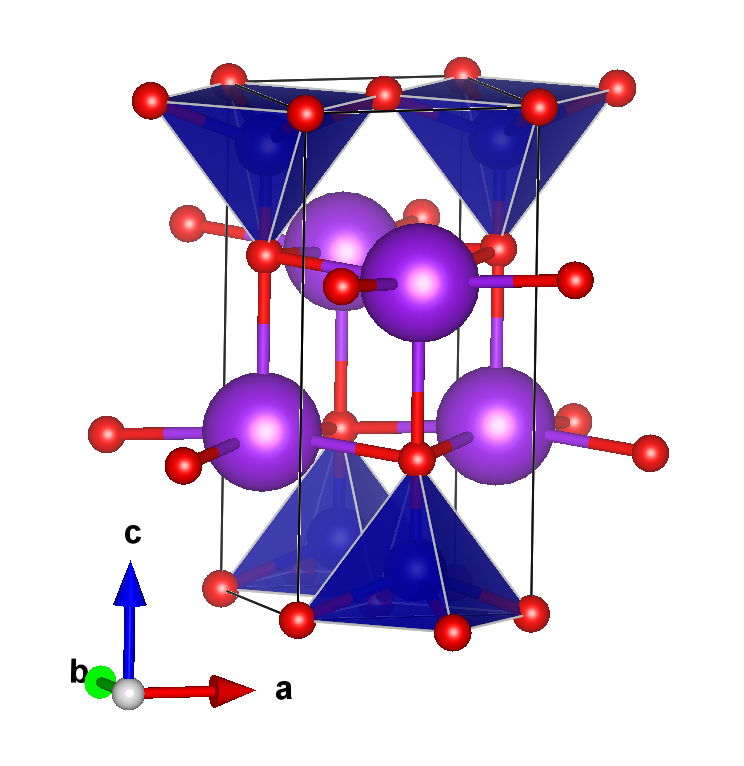}
      \includegraphics[width=4cm]{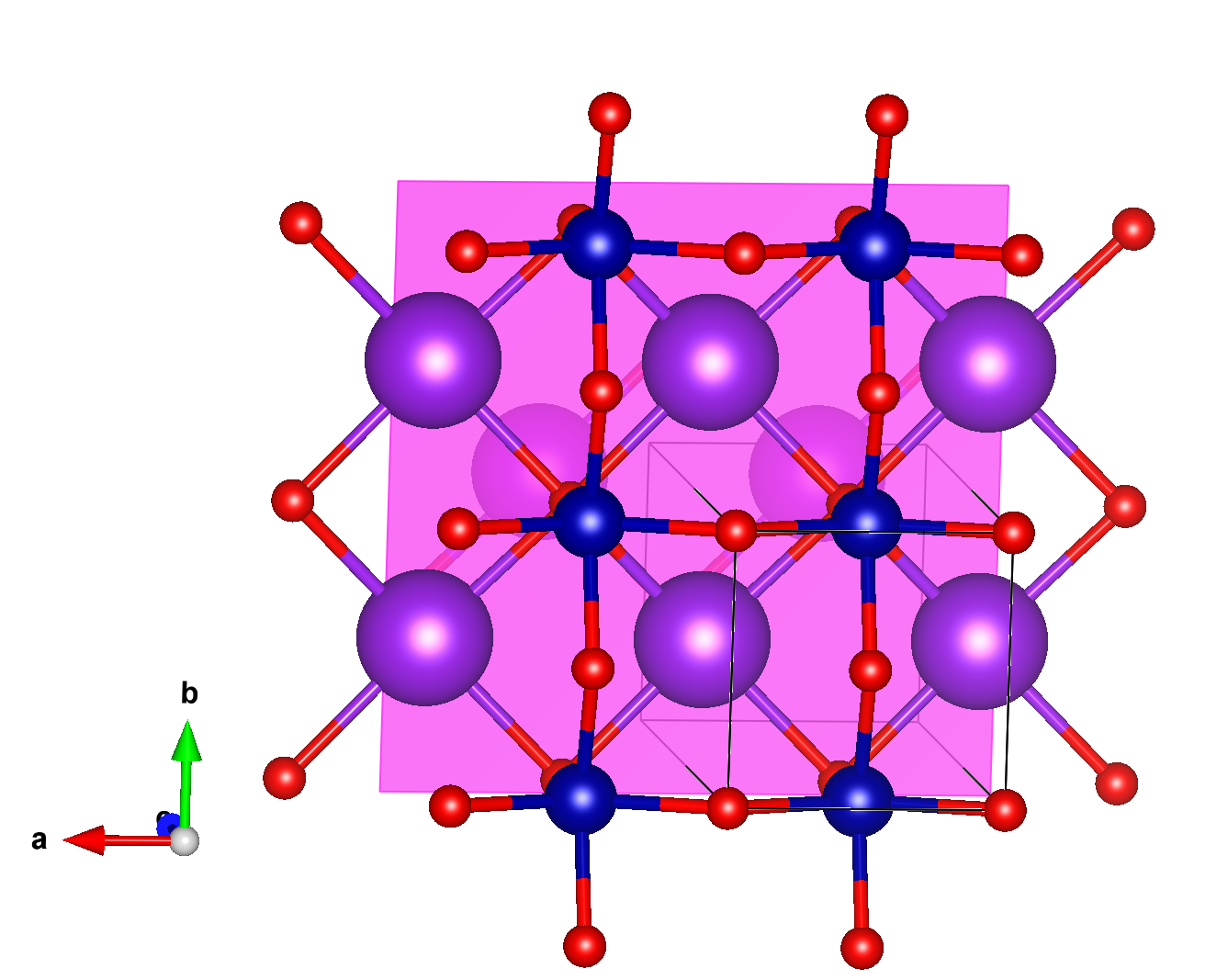}
      \caption{$P4/nmm$ crystal structure of KCoO$_2$ in two different projections, the
               first emphasizing the square pyramidal coordination of Co, and the second
               projecting on the c-plane emphasizing the Co-O square arrangement
               reminiscent of the cuprates. \label{figcryst}}
    \end{figure}

\section{Computational Methods}\label{sec:methods}
  The calculations in this study combine density functional theory (DFT) with many-body
  perturbation theory (MBPT). While DFT is used as a starting point for the electronic
  structure, the generalized gradient approximation (GGA) (used here in the
  Perdew-Burke-Ernzerhof (PBE) parametrization \cite{PBE}) is not sufficiently accurate to
  make accurate predictions for band gaps and optical properties. To calculate the
  quasiparticle band structure, we use Hedin's $GW$ method \cite{Hedin65,Hedin69} in which
  $G$ is the one-electron Green's function and  $W$ is the screened Coulomb interaction.
  More specifically, we here use the quasiparticle self-consistent version of $GW$
  (QS$GW$) \cite{Kotani07,MvSQSGWprl}, which becomes independent of the starting DFT
  approximation by including a non-local exchange correlation potential extracted from the
  $GW$ self-energy and updating the non-interacting $H_0$ Hamiltonian. By non-interacting,
  we here mean that the dynamical (energy-dependent) interactions are not included but
  only a static  interaction as in DFT.

  The band structure method used to solve the Kohn-Sham equations underlying both the DFT
  and QS$GW$ method is the full-potential linearized muffin-tin orbital (FP-LMTO) method
  as implemented in the {\sc Questaal} codes \cite{questaalpaper}. This is an augmentation
  method in which the basis set consists of atom-centered smoothed Hankel function
  spherical waves \cite{Bott98}, {\sl augmented} inside the muffin-tin spheres with
  solutions of the radial Schr\"odinger equation of the all-electron potential at a
  linearization energy and their energy derivatives. Core states, calculated with atomic
  boundary conditions at the muffin-tin radius, are thus fully included in the charge
  density (thereby including core-valence exchange) and semicore states are further
  included in the basis set as {\sl local orbitals} with a fixed boundary condition at the
  sphere radii. Here we include K $3p$ states as local orbitals. 

  In the LMTO implementation of the $GW$ method, two-point quantities such as the bare and
  screened Coulomb interaction are expanded in an auxiliary mixed product basis set, which
  incudes products of partial waves inside the spheres and interstitial plane waves. Such
  a basis set is more efficient than a plane wave basis set to describe the screening and
  reduces the need to include high-energy empty bands.

  The optical dielectric function is calculated using the Bethe-Salpeter equation (BSE) in
  the Tamm-Damcoff approximation and using a static $W$ \cite{Onida02} as implemented by
  Cunningham \etal \cite{Cunningham21,Cunningham23} in the LMTO-basis set within the
  {\sc Questaal} package.
  
  The basis set and other convergence parameters are discussed along with the results. A
  well-converged $\Gamma$-point centered ${\bf k}$-mesh of $8\times8\times4$ and the
  tetrahedron method are used for the Brillouin zone integrations in the DFT calculations.
  The atom-centered basis set allows us to interpolate the $GW$ self-energy via a Fourier
  transform to real space and back to any desired {\bf k}-point even when using a somewhat
  coarser $5\times5\times3$ mesh of points on which the $GW$ self-energy is evaluated. The
  BSE calculations are performed including 12 valence and 6 conduction bands.

  To study the magnetic exchange interactions, we use the approach of Kotani and van
  Schilfgaarde \cite{Kotani2008}, which extracts the exchange interactions from the
  transverse spin susceptibility within a rigid spin approximation within each muffin-tin
  sphere. The non-interacting spin-spin response function is first calculated from the
  spin-dependent QS$GW$ eigenstates and eigenvalues:
\begin{eqnarray}
  &&\chi_{\bf q}^{0+-}({\bf r},{\bf r}',\omega)\nonumber \\
  &&=\sum_{{\bf k}n\downarrow}^{occ}\sum_{{\bf k}'n'\uparrow}^{unocc}
  \frac{\Psi^*_{{\bf k}n\downarrow}({\bf r})\Psi_{{\bf k}'n'\uparrow}({\bf r})\Psi_{{\bf k}'n'\uparrow}^*({\bf r}')\Psi_{{\bf k}n\downarrow}({\bf r}')}{\omega-(\epsilon_{{\bf k}'n'\uparrow}-\epsilon_{{\bf k}n\downarrow})+i\delta}  \nonumber \\
    &&+\sum_{{\bf k}n\downarrow}^{unocc}\sum_{{\bf k}'n'\uparrow}^{occ}
  \frac{\Psi^*_{{\bf k}n\downarrow}({\bf r})\Psi_{{\bf k}'n'\uparrow}({\bf r})\Psi_{{\bf k}'n'\uparrow}^*({\bf r}')\Psi_{{\bf k}n\downarrow}({\bf r}')}{-\omega-(\epsilon_{{\bf k}n\downarrow}-\epsilon_{{\bf k}'n'\uparrow)}+i\delta},\nonumber \\
\end{eqnarray}
  with ${\bf k}'={\bf k}+{\bf q}$. It is then coarse-grained by averaging over the
  spheres, which constitutes the rigid spin approximation, 
\begin{equation}
D^0({\bf q},\omega)_{aa'}=\int_ad^3r\int_{a'}d^3r'\bar{e}_a({\bf r})\chi_{\bf q}^{0+-}({\bf r},{\bf r}',\omega))\bar{e}_{a'}({\bf r}'),
\end{equation}
  with $e_a({\bf r})=M_a({\bf r})/M_a$, $M_a=\int_ad^3r M_a({\bf r})$,
  $\bar{e}_a=e_a({\bf r})/\int_ad^3r|e_a({\bf r})|^2$. So the
  $\bar{e}_a({\bf r})\propto e_a({\bf r})$ is a vector along the local magnetization
  density $M_a({\bf r})=n_\uparrow({\bf r})-n_\downarrow({\bf r})$, normalized by the
  total moment per sphere $M_a$, which is normalized by
  $\int_ad^3r\bar{e_a}({\bf r})e_a({\bf r})=1$. Following Antropov \cite{Antropov2003},
  the exchange interactions defined by
\begin{equation}
  J_{\alpha\beta}({\bf r},{\bf r}')=-\frac{\delta^2E}{\delta M_\alpha({\bf r})\delta M_\beta({\bf r}')}
\end{equation}
  corresponds to the inverse of the transverse susceptibility $J=\chi^{-1}$ because the
  changes in magnetization originate from an external magnetic field and $\chi$ is the
  response function {\sl vs.} the external magnetic field. This differs from the above
  non-interacting susceptibility, which defines the response with respect to the total
  field, including the one generated by the interactions. Assuming now that the similarly
  sphere-averaged interaction term is {\bf q}-independent and site-diagonal,
  $\bar{U}_{aa'}({\bf q},\omega)=\bar{U}_a(\omega)\delta_{aa'}$ Kotani and van
  Schilfgaarde show that one can find $\bar{U}_a(\omega)$ by requiring to fulfill a
  sum rule and the $\omega\rightarrow\infty$ asymptotic behavior. This then yields
  directly the inverse of the interacting $D({\bf q},\omega)$ as
\begin{eqnarray}
 [D({\bf q},\omega)]^{-1}_{aa'}&=&\frac{\omega}{M_a}\delta_{aa'}-\bar{J}_{aa'}({\bf q},\omega),\nonumber\\
 \bar{J}_{aa'}({\bf q},\omega)&=&-[D^0({\bf q},\omega)]^{-1}_{aa'}\nonumber \\
 &+&\left(\sum_b M_b[D^0({\bf q}=0,\omega)]^{-1}_{ba}/M_a\right)\delta_{aa'}.\nonumber \\
\end{eqnarray}
  This $\bar{J}_{aa'}$ is closely related to the Heisenberg exchange interactions, as
  discussed further in \cite{Kotani2008},
\begin{equation}
  [D^{\cal H}({\bf q},\omega)]^{-1}_{aa'}=\frac{\omega}{M_a}\delta_{aa'}-J^{\cal H}({\bf q}),
\end{equation}
  which suggest taking $J^{\cal H}({\bf q})=J({\bf q},\omega=0)_{aa'}$, \ie the static
  limit. Here $J_{aa'}$ differs from $\bar{J}_{aa'}$ only by removing the on-site term of
  $[D^0]^{-1}$. Although, we sketched here the more general presentation of
  \cite{Kotani2008} of the enhanced susceptibility, in the end, we obtain the exchange
  interactions from the static version of the inverse of the bare susceptibility,
  $[D^0]^{-1}$. This is also compatible with the Liechtenstein \etal multiple scattering
  formulation of the linear response theory \cite{Liechtenstein87}.

  The Heisenberg exchange interactions in real space $J_{a,a'}^{0{\bf T}}$ can then be
  obtained by inverting the Bloch sum
  $J_{aa'}({\bf q})=\sum_{\bf T} e^{i{\bf q}\cdot{\bf T}}J_{a,a'}^{0{\bf T}}$, \ie by an
  integral over the Brillouin zone. Thus, if we calculate the $J_{aa'}({\bf q})$ on a
  $N\times N\times N$ mesh in the Brillouin zone, then a discrete inverse Fourier
  transform gives us the $J_{a,a}^{0{\bf T}}$ in a $N\times N\times N$ supercell, or the
  exchange interactions out to a distance $|{\bm \tau}_{a'}-{\bm \tau}_a+{\bf T}_{max}|$
  where ${\bm \tau}_a$ is the position in the unit cell of the atom labeled $a$.
  ${\bf T}_{max}$ is the largest lattice translation vector corresponding to the
  superlattice.

  We can then use various methods to estimate the critical temperature $T_c$, such as the
  mean-field approximation, or the random phase approximation (RPA) developed by Tyablikov
  \etal\cite{Tyablikov1959} and Callen \cite{Callen63} and more recently used by Rusz
  \etal\cite{Rusz2005}. In the case (which occurs here) of several magnetic sites per unit
  cell, the critical temperature in the mean-field approximation is obtained by
  diagonalizing the matrix of the $J^0_{ab}=\sum_{\bf T} J_{ab}^{0{\bf T}}$, where if
  $a=b$, the on-site term $J_{aa}^{00}$ is excluded from the sum,
\begin{equation}
  \sum_b[J_{ab}^0-j\delta_{ab}]\langle S_b\rangle=0.
\end{equation}
  The mean-field critical temperature is then given by $k_BT_c=2j_{max}/3$ with $j_{max}$
  the highest eigenvalue \cite{Sasioglu2004}. The normalized eigenvectors tell us the
  relation of the average spins on the sites in the cell, \ie the type of magnetic
  ordering. The mean-field critical temperature is used as starting point for the RPA
  iterative procedure. Typically, the latter gives an underestimate while the mean-field
  method gives an upper limit. 
 
\section{Results}
\subsection{Band structure and magnetic moments in ferromagnetic structure}
    \begin{figure}
      \includegraphics[width=8cm]{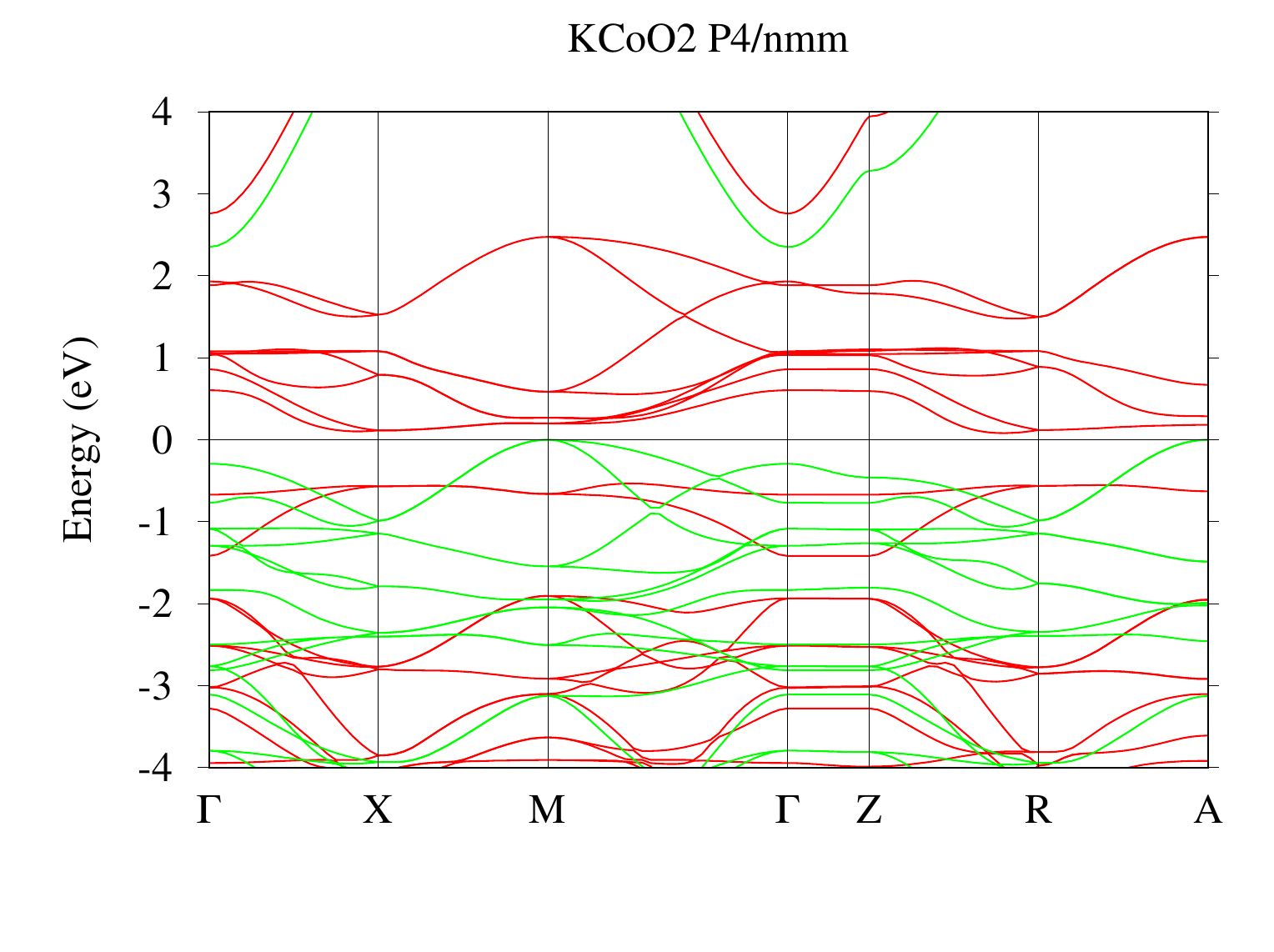}
      \caption{Spin-polarized band structure in GGA of KCoO$_2$ in the $P4/nmm$ phase.
               \label{figbnds}}
    \end{figure}

    \begin{figure}
      \includegraphics[width=8cm]{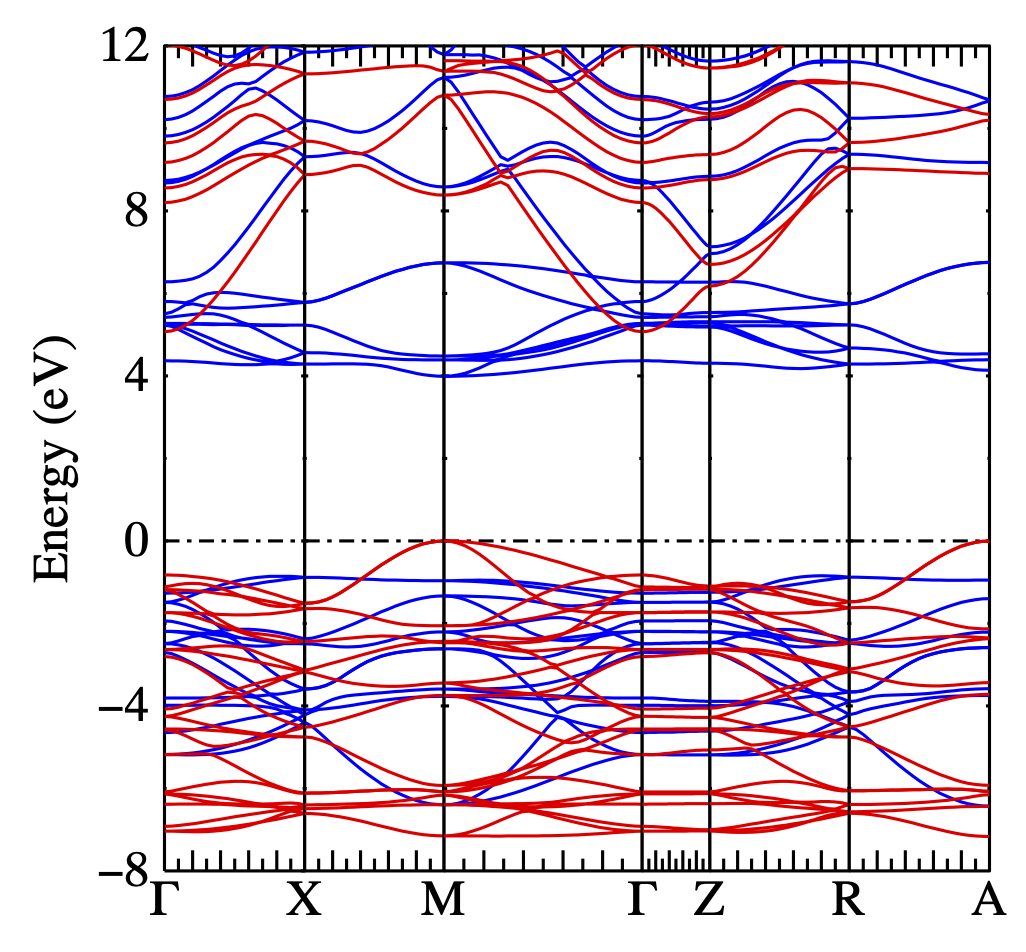}
      \caption{Band structure of KCoO$_2$ in QS$GW$ approximation for the $P4/nmm$
               structure.\label{figbndgw}}
    \end{figure}

  The crystal structure is shown in Fig.~\ref{figcryst}. There are two types of O, the
  O$_{(1)}$ lying close to the K {\bf c}-plane, which are strongly bonded to the Co in the
  $z$-direction at a bond distance of 1.741 \AA, and the O$_{(2)}$ which lie in the
  Co-O$_{(2)}$ layer and have a bond-length to Co of 2.063 \AA. The K-O$_1$
  in-{\bf c}-plane bond length is 2.732 \AA\  and along the ${\bf c}$-axis is 2.791 \AA.
  Within spin-polarized GGA, we find a high magnetic moment of 4 $\mu_B$ per Co atom and a
  ferromagnetic semiconductor band structure with a small gap. The high magnetic moment
  can be understood as follows. Within the square pyramidal coordination of Co and
  choosing $x$ and $y$ axis pointing toward the oxygen neighbors, the $x^2-y^2$ orbitals
  have a large $\sigma$-type antibonding interaction with O-p$_x$ and $p_y$ orbitals. In
  the $D_4$ point group they correspond to the $b_1$ irreducible representation. The
  $3z^2-r^2$ orbital ($a_1$) also has a strong interaction with O $p_z$ on one side but
  opposite to it lies a K ion which electrostatically would tend to pull this level down
  in terms of crystal field splitting. The $xy$ ($b_2$) has the weakest interaction, while
  the $xz,yz$ doubly degenerate $e$ representation has intermediate $\pi$-like
  interaction. For $d^6$ we can thus expect the configuration
  $b_2^2e_\uparrow^2a_{1\uparrow}^1b_{1\uparrow}^1$ where the spin-polarization of the
  degenerate $e$ level promotes the exchange splitting of the higher $a_1$ and $b_1$ level
  becoming larger than the crystal field splitting.

    \begin{table}
      \caption{Band gaps in $P4/nmm$ KCoO$_2$ in QS$GW$ approximation.\label{tabgaps}}
      \begin{ruledtabular}
        \begin{tabular}{lcccc}
          type    & $k_v$    & $k_c$    & spin type              & gap (eV) \\ \hline
         direct   & M        & M        & $\uparrow\downarrow$   & 3.99     \\
         indirect & M        & $\Gamma$ & $\uparrow\uparrow$     & 5.07     \\
         direct   & M        & M        & $\uparrow\uparrow$     & 8.38     \\
         direct   & $\Gamma$ & $\Gamma$ & $\uparrow\uparrow$     & 5.90     \\
         indirect & 0.8Z-R   & M        & $\downarrow\downarrow$ & 4.83     \\
         direct   & M        & M        & $\downarrow\downarrow$ & 4.95     \\
         direct   & 0.8Z-R   & 0.8Z-R   & $\downarrow\downarrow$ & 5.05
        \end{tabular}
      \end{ruledtabular}
    \end{table}
    
    \begin{figure}
      \includegraphics[width=9cm]{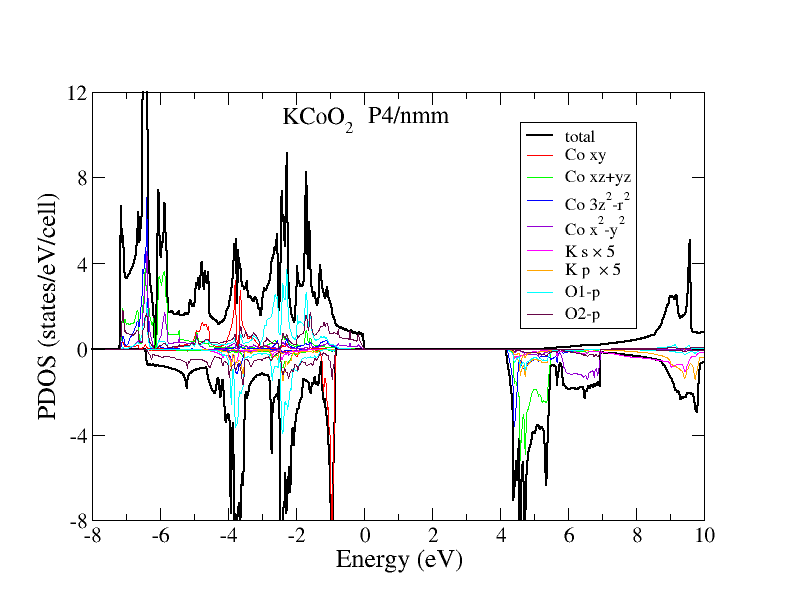}
        \caption{Total and partial densities of states for KCoO$_2$ in $P4/nmm$ structure.
                 \label{figpdos}}
    \end{figure}
    
    \begin{figure*}[h]
      \includegraphics[width=8cm]{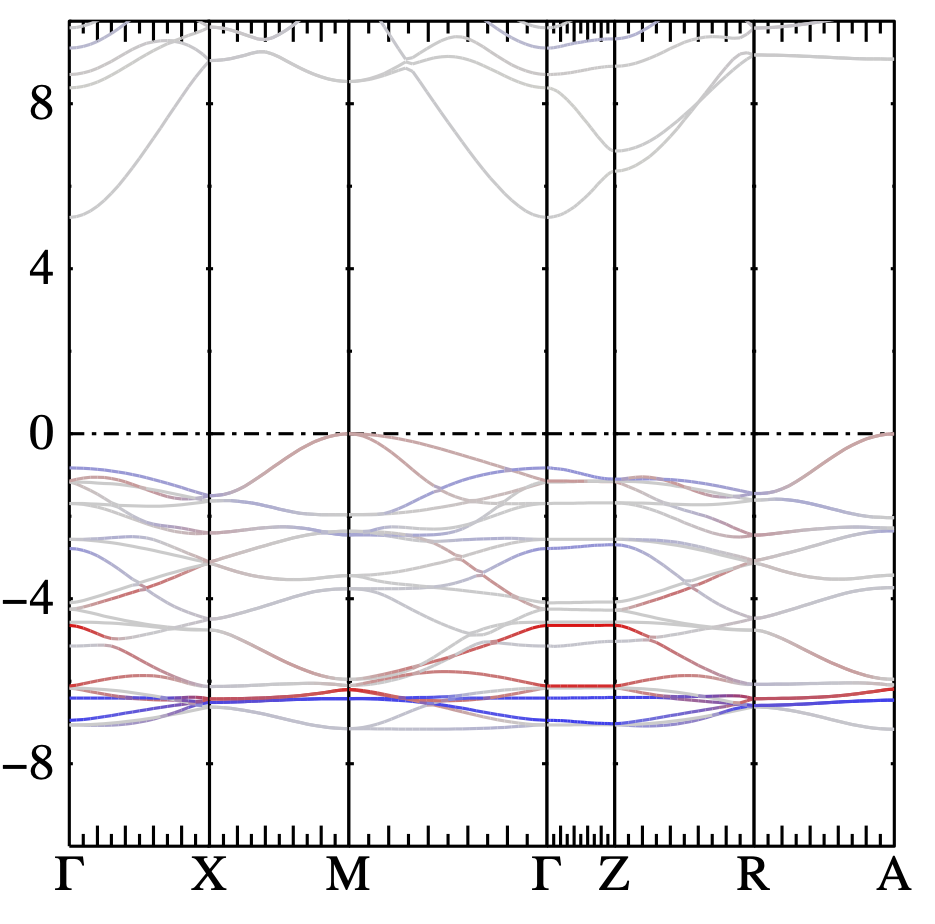}
      \includegraphics[width=8cm]{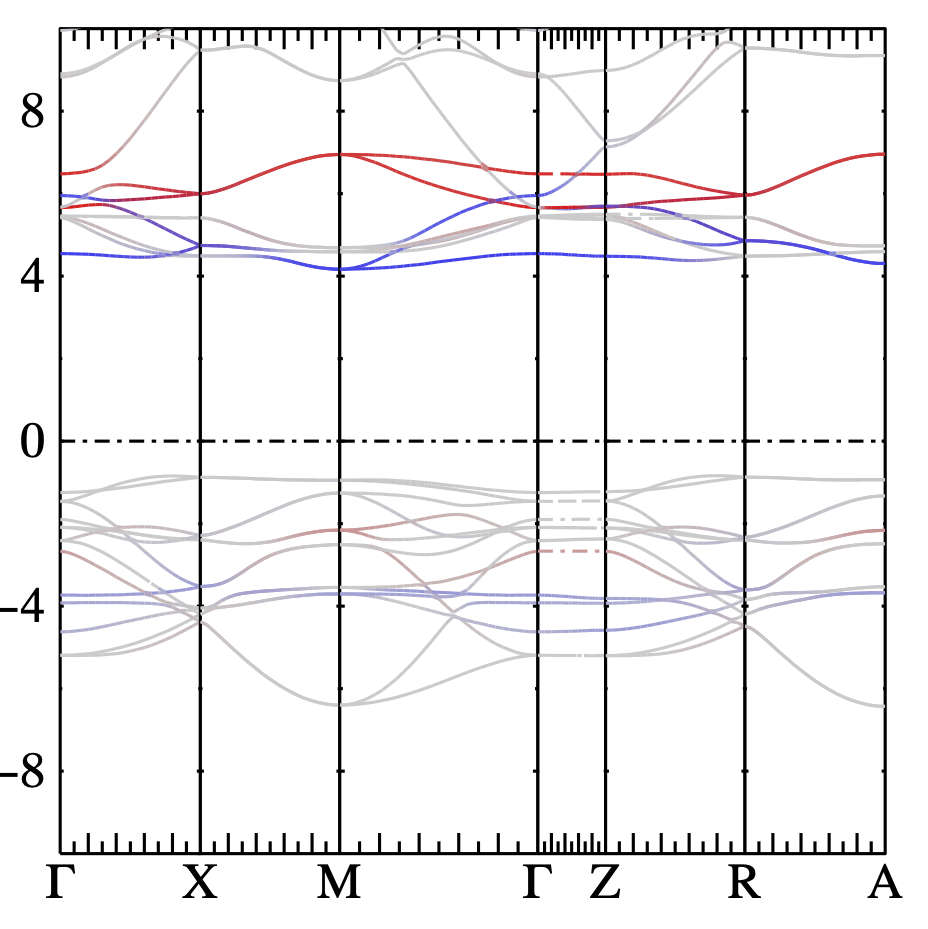}
      \includegraphics[width=8cm]{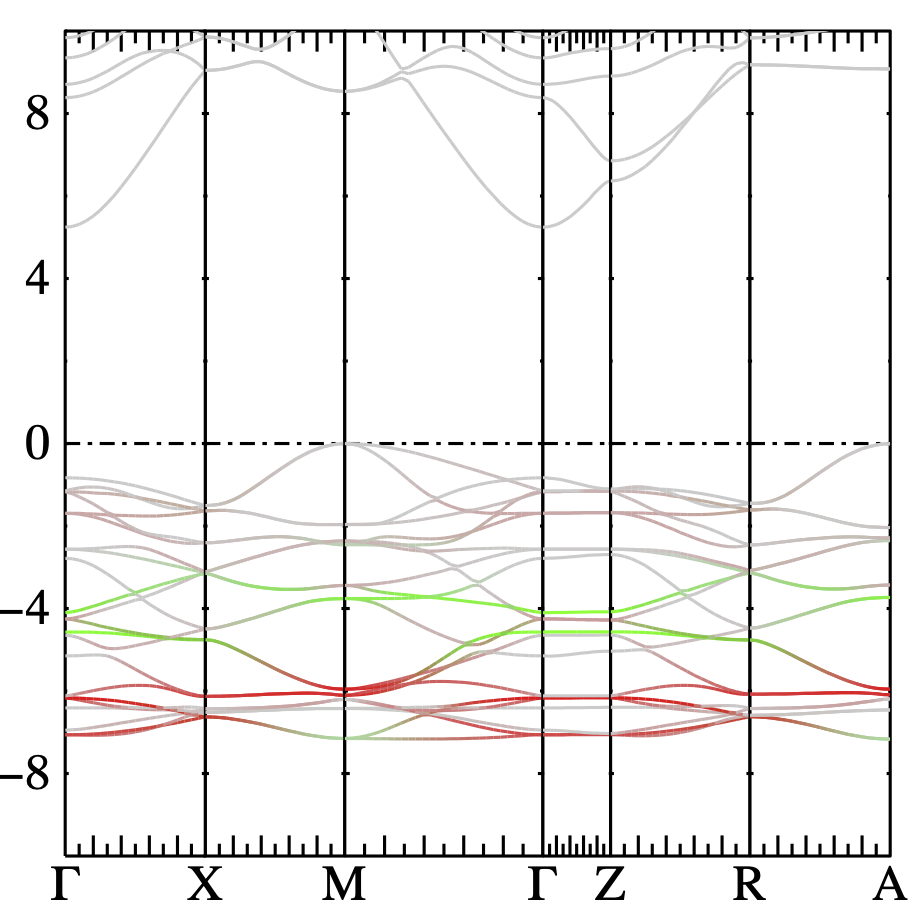}
      \includegraphics[width=8cm]{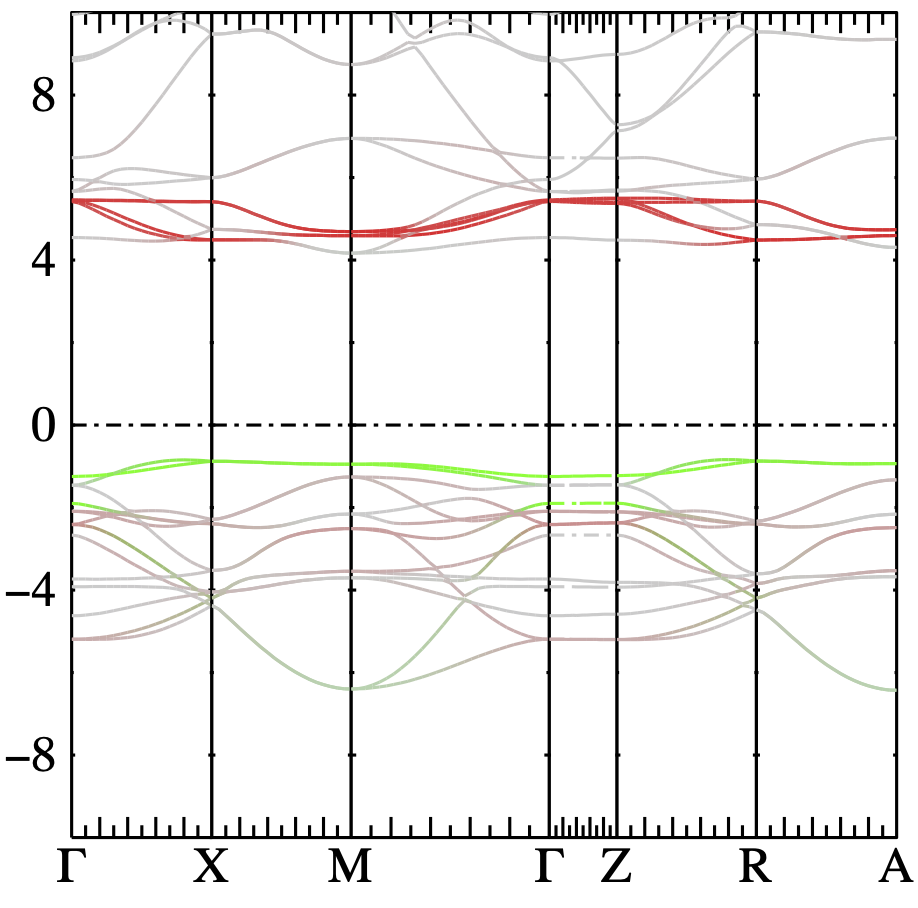}
      \caption{Orbital weights of Co $d$ of the bands, left for majority and right for
               minority spin; top row (red) $x^2-y^2$, (blue) $3z^2-r^2$,  bottom row
               (red) xz,yz and green (xy).\label{figbndcol}}
    \end{figure*}
    
    \begin{figure*}
      \includegraphics[width=8cm]{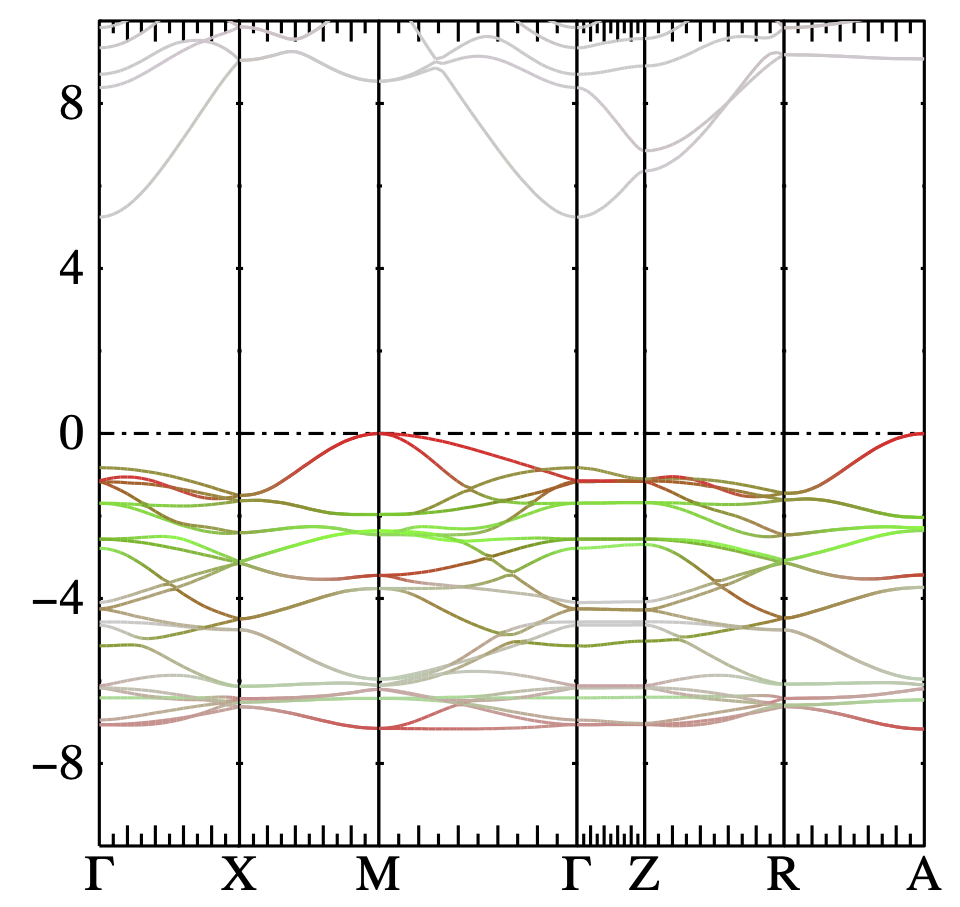}
      \includegraphics[width=8cm]{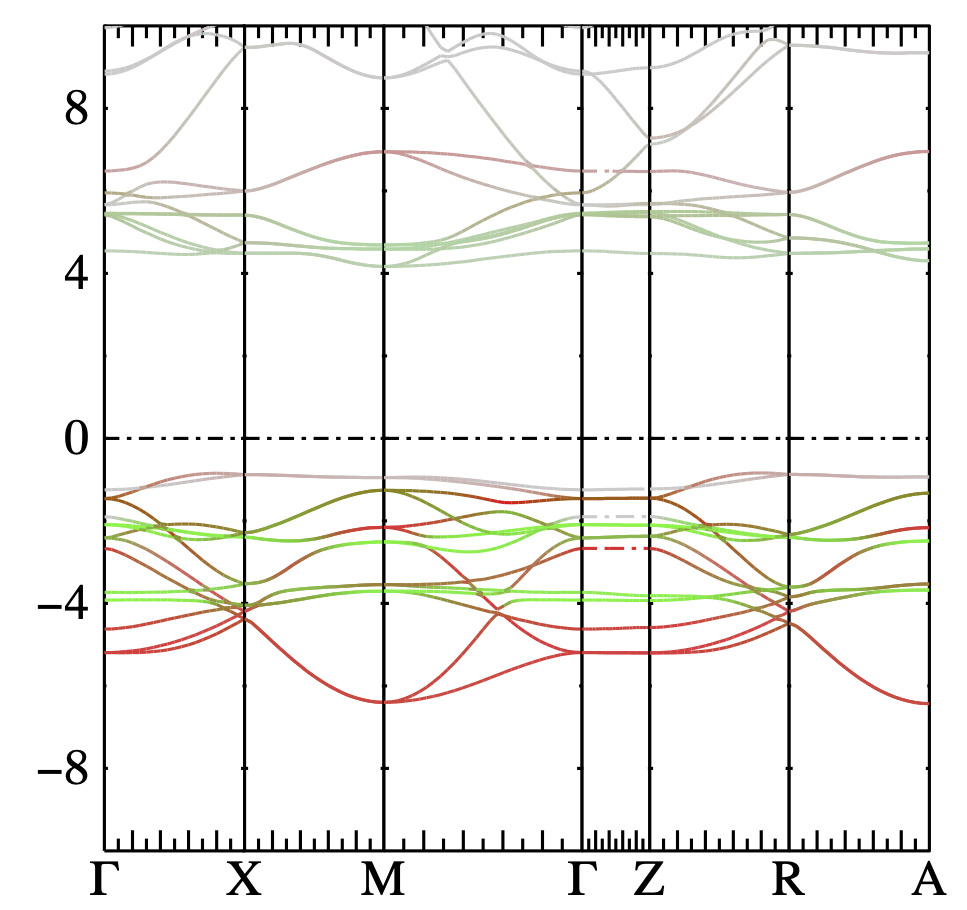}
      \caption{Orbital weights of O $p$ of the bands for left (majority) and right
               minority spin; red(O$_2$), green (O$_1$)\label{figbndocol}} 
    \end{figure*}
    
    \begin{figure*}
      \includegraphics[width=8cm]{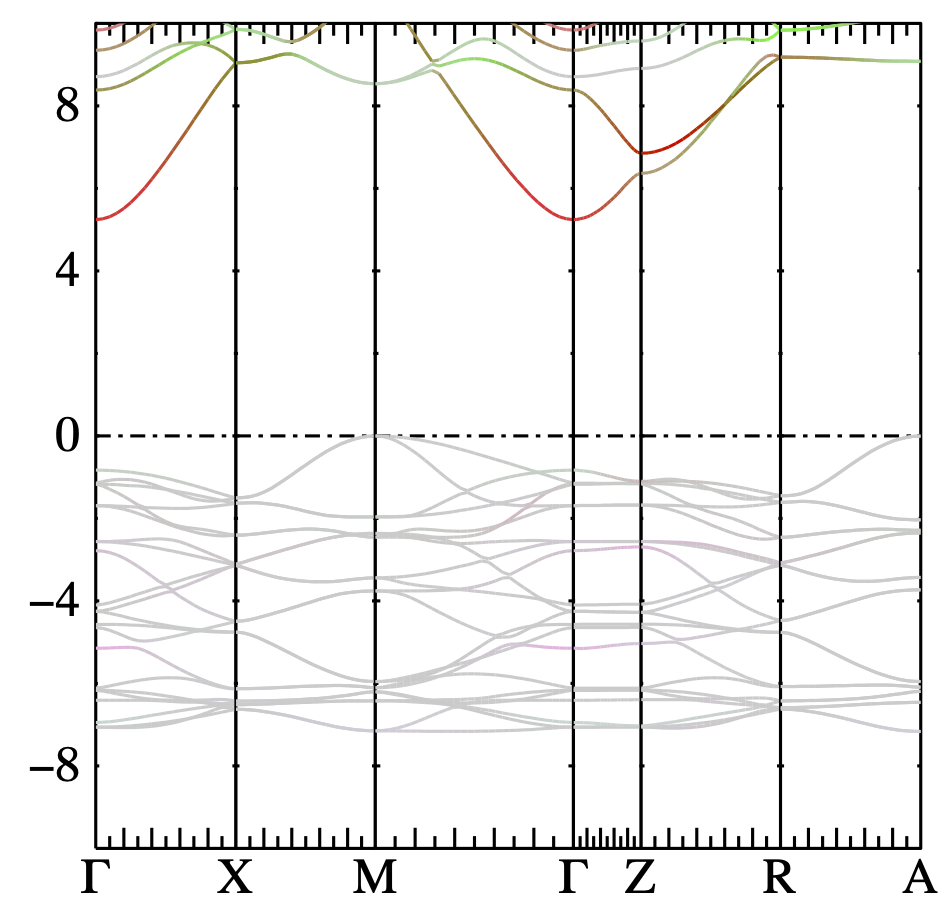}
      \includegraphics[width=8cm]{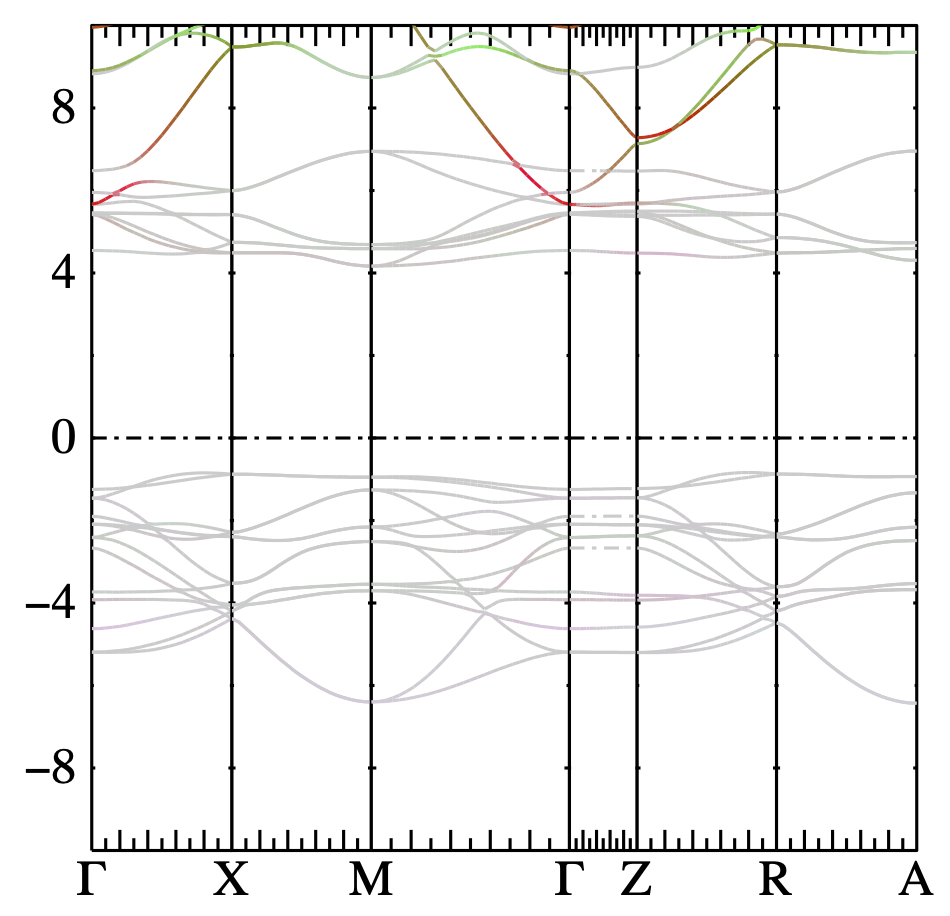}
      \caption{Orbital weights of K $s$ (red) and $p$ (green) for left (majority) and
               right (minority) spin.\label{figbndKcol}}
    \end{figure*}

  The spin-polarized band structure at the GGA level is shown in Fig.~\ref{figbnds}.
  Interestingly, a small gap opens between spin-up and spin-down bands with rather flat
  bands. The gap becomes significantly larger in the QS$GW$ method as seen in
  Fig.~\ref{figbndgw}. We can see that the VBM occurs at $M$ for the majority spin while
  the CBM has minority spin character. The reason why the VBM occurs at $M$ is that the
  antibonding interaction of the $x^2-y^2$ orbitals with O-$p\sigma$ orbitals is
  optimized at this {\bf k}-point by the Bloch function phase factors because the same
  sign lobes point toward the Co for each O along in the square coordination of Co. The
  majority spin CBM occurs at $\Gamma$. The majority spin highest valence band is quite
  flat. The majority spin and minority spin band gaps are staggered with respect to each
  other. The relevant gaps are summarized in Table~\ref{tabgaps}. These calculations used
  a $2\kappa$ LMTO basis set with smoothed Hankel function envelope functions up to $l=2$
  ($spdspd$) for K, and O and ($spdfspd$) for Co. Without the Co-$f$ the gap is slightly
  higher (4.17 eV).

  The total and partial densities of states (PDOS) on various orbitals are shown in
  Fig.~\ref{figpdos}. The orbital contributions of the bands are shown in
  Figs.~\ref{figbndcol},\ref{figbndocol} and \ref{figbndKcol}. The PDOS refers to a
  partial wave decomposition while the bands correspond to a decomposition in
  muffin-tin-orbital basis functions. These results were obtained with the slightly
  smaller basis set without the Co-$f$ basis functions but for the qualitative
  features, this is of no importance. The PDOS and colored band plots provide consistent
  information.

  We can see that the bands with predominantly $xy$ character are filled for both spins.
  On the other hand, the minority spin $x^2-y^2$, $3z^2-r^2$ and $xz,yz$ contributions
  occur mostly in the empty bands. This is consistent with the above described origin of
  the large magnetic moment of $4\mu_B$. The valence bands also have a significant
  contribution from O-$p$ as shown in Fig.~\ref{figbndocol}. The K-contributions (shown
  in Fig.~\ref{figbndKcol}), as expected, occur mostly in the conduction band. They do
  not contribute significantly to the lower lying set of minority spin bands. This is
  consistent with K donating electron to the CoO$_2$ layer.

  One can see that the majority spin VBM in terms of Co-$d$ has mostly $x^2-y^2$
  contribution but its dominant character is O$_{(2)}$-$p$. In other words, it is an
  antibonding state between O$_{(2)}-p$ in the CoO$_2$ layer and Co-$d$-$x^2-y^2$ orbital.
  The conduction band minimum which nominally occurs at $M$ but corresponds to a rather
  flat band has predominantly minority spin $3z^2-r^2$ character on Co and a much smaller
  O-$p$ character. Thus direct optical transitions from the VBM to the CBM are charge
  transfer type but would only be allowed for circularly polarized light because they are
  from spin-up to spin-down. For linearly polarized light the optical transitions would be
  mostly between the minority spin bands which are both quite flat and have both Co-$d$
  character, however with $xy$ character for the VBM of minority spin and $3z^2-r^2$
  character for the conduction band minimum. This would require a change of $m$ orbital
  character of $\Delta m=2$ and is therefore forbidden in the electric dipole
  approximation. On the other hand for majority spin, the transitions would be indirect
  and therefore also forbidden. These optical properties are rather unique and intriguing.

\subsection{Optical dielectric function in FM state.}
    \begin{figure}
      \includegraphics[width=8cm]{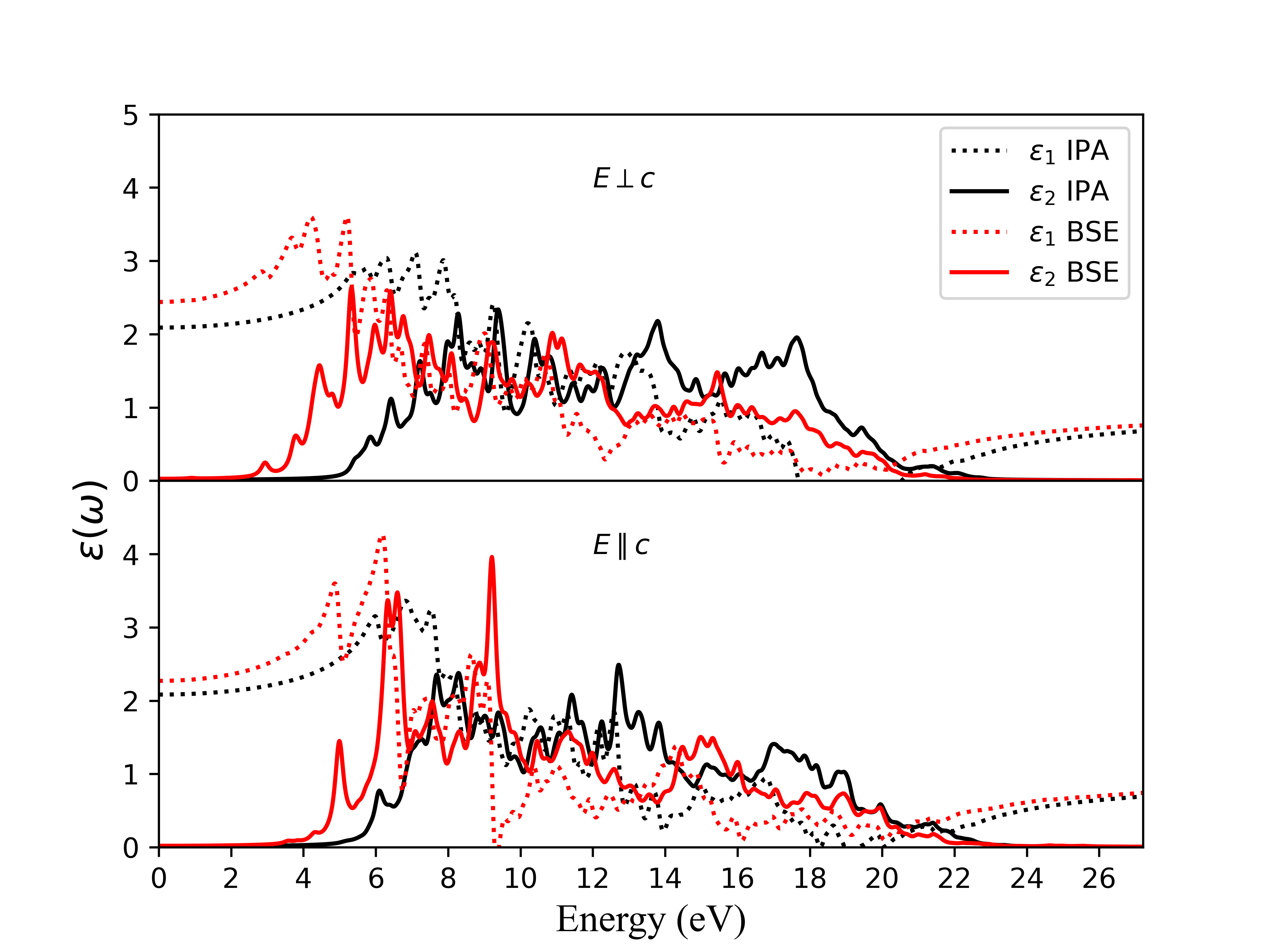}
      \caption{Optical dielectric function for two polarizations in IPA and BSE, assuming
               transitions only between the equal spins.\label{figeps}}
    \end{figure}

  The optical dielectric function was calculated assuming only transitions between
  majority spin to majority spin states, and minority to minority spins. They were
  calculated in the independent particle approximation (IPA) and using the Bethe-Salpeter
  equation (BSE), which includes local field and electron-hole interaction effects. The
  results are shown for both the real and imaginary part in Fig.~\ref{figeps}. Within both
  IPA and BSE we assume here that the allowed dipole transitions are spin separated.
  Strictly speaking, the exchange Coulomb interaction matrix elements in the BSE
\begin{equation}
\begin{split}
  &V_{vc{\bf k}v'c'{\bf k}'}= \\&\int d(1) d(2) \psi_{v{\bf k}}(1)\psi_{c{\bf k}}(1)^*
  v(|{\bf r}-{\bf r}'|)\psi_{v'{\bf k}'}(2)^*\psi_{c'{\bf k}'}(2)
\end{split}
\end{equation}
  require the valence and conduction band at one {\bf k} to have the same spin and also at
  the other {\bf k}-point because the integral over coordinates (1) or (2) includes spin
  summation, but the spins of (1) and (2) may differ. So, an interaction between up and
  down spin is mediated by the exciton exchange interaction. However, if the optical
  transitions of the separated spins are sufficiently well separated, we may ignore this
  interaction. This is the approximation we currently are making. Note that for
  non-spin-polarized systems, this is manifestly not the case since the up-up and
  down-down transitions are degenerate. However, in that case the spin structure of the
  excitons is clear \cite{Rohlfing2000} and leads to dark spin triplets involving only the
  $-W$ matrix elements with the screened Coulomb interaction, whereas the spin-singlet
  ones involve $2\bar{V}-W$ with $\bar{V}$ the microscopic part of the above defined
  exciton exchange interactions (\ie excluding the average or long-range macroscopic
  interaction). At present we include $\bar{V}-W$ but separately for each spin. In the
  calculation in Fig.~\ref{figeps} we include 24 valence bands and 16 conduction bands.
  This includes the 12 O-$2p$ bands and 10 Co-$d$ states of each spin.  

  With this understanding of the approximations made, we now examine the results. First,
  we may note that the electron-hole effects have a significant impact with excitonic
  peaks occurring below the quasiparticle gap. Interestingly, there seems to be almost a
  uniform red shift of $\sim$2 eV from IPA to BSE. Next, we caution that the
  optical matrix elements of the velocity matrix elements may be overestimated because of
  difficulties in evaluating the contributions from the non-local self-energy operator of
  the $GW$ approximation, which require estimating the $d\Sigma/dk$ numerically. This has
  been found in other systems to overestimate the magnitude of $\varepsilon_2(\omega)$
  compared to evaluating the $\varepsilon_2({\bf q},\omega)$ at finite small ${\bf q}$ and
  then extrapolating to ${\bf q}\rightarrow0$ numerically. This then also leads to an
  overestimate of the $\varepsilon_1(\omega)$ below the gap and in particular its limit
  $\lim_{\omega\rightarrow0}\varepsilon_1(\omega)$, which gives the electronic screening
  at the static limit. In other words, this ``static limit'' does not include phonon
  contributions (conventionally referred to as $\varepsilon_\infty$) and corresponds to
  the square of the index of refraction $n^2$ in the range sufficiently well below the gap
  but higher than any of the phonon modes. Also, at present we cannot yet calculate the
  transitions between up and down spin bands which would be of great interest in this
  system but are expected to occur only for circularly polarized light. 

\subsection{Magnetic ordering}
  Having established the formation of large magnetic moments as a basic way to stabilize
  the $d^6$ configuration in the unusual pyramidal environment, we now turn to the
  question of their ordering.

    \begin{table}
      \caption{Exchange interactions in mRy; $a$ and $b$ label the magnetic atoms in the
               cell, ${\bf T}$ gives the lattice vector in reduced coordinates, $z$ the
               number of equivalent neighbors in the star, $J_{ab}^{0{\bf T}}$, he
               exchange interaction in mRy and the last column gives the cumulative sum,
               with the last row for a given block of $a,b$ giving the cumulative sum up
               to $|{\bm\tau}_a-{\bm \tau}_b-{\bf T}|<r_{cut}=5$ in units of the in-plane
               tetragonal lattice constant $a$.\label{tab:exchange}}
      \begin{ruledtabular}
      \begin{tabular}{llccrc}\\
        a & b & ${\bf T}$ & $z$ & $J_{ab}^{0{\bf T}}$ & $\sum_{\bf T}J_{ab}^{0{\bf T}}$ \\
      \hline
        1 & 1 & $(1,0,0)$ & 4   & $-$0.3725           & $-$1.490 \\
        1 & 1 & $(1,1,0)$ & 4   & $-$0.0362           & $-$1.635 \\
        1 & 1 & $(2,0,0)$ & 4   &    0.0012           & $-1.630$ \\
        1 & 1 & \dots     &     &                     & $-1.561$ \\
      \hline
        1 & 2 & $(0,0,0)$ & 4   & $-$0.0342           & $-0.137$ \\
        1 & 2 & $(1,0,0)$ & 8   &    0.0024           & $-0.118$ \\
        1 & 2 & $(0,0,1)$ & 4   & $-$0.0024           & $-0.127$ \\
        1 & 2 & \dots     &     &                     & $-0.137$ \\ 
      \end{tabular}
      \end{ruledtabular}
    \end{table}
  We start from the ferromagnetic unit cell and use the procedure outlined in
  Sec.~\ref{sec:methods}. Table~\ref{tab:exchange} shows some of the near neighbor
  exchange interactions and their cumulative sums. The $J_{ab}^0$ matrix has the form
\begin{equation}
  J^0=\left(\begin{array}{cc} J_{11}^0 & J_{12}^0  \\
                              J_{12}^0 & J_{11}^0\end{array}\right)
\end{equation}
  with eigenvalues $J_{11}^0\pm J_{12}^0$, hence we find a mean-field
  $T_c=(2/3k_B)(J_{11}^0-J_{12}^0)$ of $-149$ K. The negative value indicates that the
  system actually wants to order antiferromagnetically. Indeed, we find that the average
  spins on site 1 and 2 are opposite for the eigenvalue $J_{11}^0-J_{12}^0$. However, we
  also see that the atoms in (100) neighboring cells also have a negative exchange
  interaction, which is in fact an order of magnitude larger. Thus, the system would
  prefer AFM ordering along the [110] direction, with parallel spins in (110) planes and
  alternating up and down spin from plane to plane but also have the two atoms inside the
  cell with opposite spin.

    \begin{figure}
      \includegraphics[width=8cm]{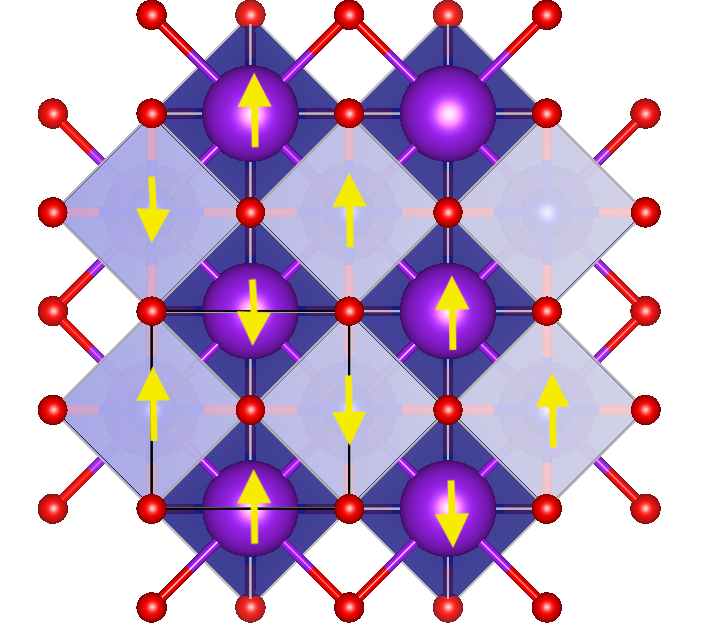}
      \caption{Spin arrangement on the Co sites, the dark blue squares indicate the Co
               pyramid of Co slightly below the plane and the light blue ones the Co
               slightly above the plane. The purple spheres are the K on top of the Co.
               The red spheres are the O and the yellow arrows indicate the spins on Co.
               \label{figspin}}
    \end{figure}

  The spin-arrangement is illustrated in Fig.~\ref{figspin}. Within the RPA method, we
  obtain a slightly lower critical temperature of $131$ K. Given that the dominant
  exchange interaction corresponds to two Co atoms in line with an O in between, we can
  interpret this as an antiferromagnetic super-exchange interaction, which might be
  dominated by the Co-$d$ $x^2-y^2$ $\sigma$-bonds with O-$p$ orbitals along the line. We
  can also see that the exchange interactions rapidly decrease beyond the first few
  neighbors. 

  \begin{figure}
    \includegraphics[width=8cm]{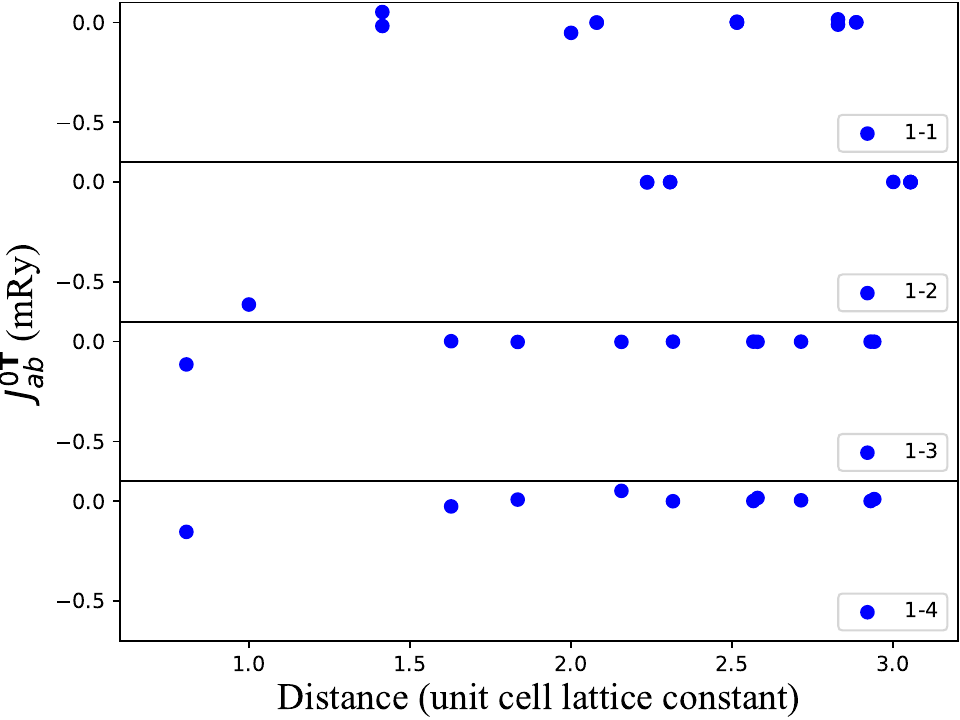}
    \caption{Exchange interactions in AFM case  as function of distance.\label{figJafm}}
  \end{figure}

  Based on the above prediction of antiferromagnetic ordering, we then construct a doubled
  cell rotated by 45$^\circ$ and recalculate the exchange interactions based on this AFM
  reference state. We calculated the total energies in the GGA while adding the energy
  independent self-energy matrix to the one-particle Hamiltonian to have the correct gaps.
  This gives $E_{AFM}-E_{FM}=-0.240$ eV/Co atom. So, the AFM ordering is definitely
  preferable. Setting $E_{AFM}-E_{FM}=2zJ_{eff}$ with $z=4$ the number of neighbors, this
  corresponds to an effective  AFM exchange interaction $J_{eff}=-30$ meV. In the
  mean-field approximation $T_N=2J_{eff}/3k_B$  and we find $T_N\approx232$ K. However,
  this assumes there is only an effective nearest neighbor interaction. 

  Next, we calculate again the exchange interactions from the magnetic susceptibility for
  this case. We now have four magnetic sites per cell. The exchange interactions between
  near neighbors are given in Table~\ref{tab:Jafm} and plotted as function of distance in
  Fig.~\ref{figJafm}.
    \begin{table}
      \caption{Exchange interactions in [110] AFM cell; magnetic atom labels $a,b$,
               lattice vector (in Cartesian coordinates and units of $a$ in $x,y$
               directions and $c$ in $z$-direction) of the $P4/nmm$ primitive cell, number
               of equivalent atoms in the star, $J_{ab}^{0{\bf T}}$ in mRy and their
               cumulative sum.\label{tab:Jafm}}
    \begin{ruledtabular}
    \begin{tabular}{llccrc}\\
      a & b & ${\bf T}$     & $z$ & $J_{ab}^{0{\bf T}}$ & $\sum_{\bf T}J_{ab}^{0{\bf T}}$ \\ \hline
      1 & 1 & $(-1,1,0)$    & 2   &    0.0525           &    0.105 \\
      1 & 1 & $((1,1,0)$    & 2   & $-$0.0174           &    0.070 \\
      1 & 1 & $(0,2,0)$     & 4   & $-0.0518$           & $-$0.137 \\
      1 & 1 & $(0,0,\pm1)$  & 2   & $-0.0002$           & $-$0.138 \\
      1 & 1 & $(-1,1,\pm1)$ & 4   & 0.0035              & $-$0.124 \\
      1 & 1 & $(1,1,\pm1)$  & 4   & $-$0.0003           & $-$0.125 \\
      1 & 1 & $(2,2,0)$     & 2   & 0.016               & $-$0.093 \\
      1 & 1 & $(-2,2,0)$    & 2   & $-$-0.0107          & $-$0.111 \\
      1 & 1 & $(2,0,\pm1)$  & 8   & $0.0009$            & $-$0.107 \\ \hline
      1 & 2 & $(1,0,0)$     & 4   & $-$0.6139           & $-$2.456 \\
      1 & 2 & $(1,2,0)$     & 4   & $-$0.0019           & $-$2.463 \\
      1 & 2 & $(-2,1,0)$    & 4   & $-$0.0006           & $-$2.466 \\
      1 & 2 & $(1,0,1)$     & 8   & $-$0.0002           & $-$2.467 \\
      1 & 2 & $(0,3,0)$     & 4   &    0.0006           & $-$2.465 \\
      1 & 2 & \dots         &     &                     & $-$2.464 \\ \hline
      1 & 3 & $(0,0,0)$     & 2   & $-$0.1144           & $-$0.229 \\
      1 & 3 & $(0,1,0)$     & 4   & 0.0020              & $-$0.221 \\
      1 & 3 & $(0,0,-1)$    & 2   & $-$0.0016           & $-$0.224 \\
      1 & 3 & $(1,1,0)$     & 2   & $-$0.0012           & $-$0.226 \\
      1 & 3 & $(0.1,-1)$    & 4   & $-$0.0010           & $-$0.227 \\
      1 & 3 & \dots         &     &                     & $-$0.231 \\ \hline
      1 & 4 & $(0,0,0)$     & 2   & $-$0.1538           & $-$0.308 \\
      1 & 4 & $(0,1,0)$     & 4   & $-$0.0263           & $-$0.413 \\
      1 & 4 & $(0,0,-1)$    & 2   &    0.0080           & $-$0.397 \\
      1 & 4 & $(-1,1,0)$    & 2   &    0.0518           & $-$0.293 \\
      1 & 4 & $(0,1,-1)$    & 4   & $-$0.0003           & $-$0.294 \\
      1 & 4 & $(0,0,1)$     & 2   &    0.0009           & $-$0.292 \\
      1 & 4 & $(0,2,0)$     & 4   &    0.0162           & $-$0.228 \\
      1 & 4 & $(-1,-1,-1)$  & 2   &    0.0044           & $-$0.219 \\
      1 & 4 & $(0,1,1)$     & 4   &    0.0008           & $-$0.216 \\
      1 & 4 & $(2,1,0)$     & 4   &    0.0106           & $-$0.173 \\
      \end{tabular}
      \end{ruledtabular}
    \end{table}

  We may note that the cumulative sum only slowly converges. The nearest neighbor
  interactions between atoms 1 and 2 in the AFM cell, corresponds to the interaction
  between atoms 1 in neighboring cells in the (100) direction in the FM cell, and is seen
  to be the dominant interaction, which is antiferromagnetic. Its value $-0.6139$ is
  almost twice as large as when we started from the FM reference state, $-0.3725$. The
  nearest interactions between 1 and 3 correspond to the interaction between the Co
  originally in the same FM cell and between Co slightly above and slightly below the
  plane. Its value is $-0.1144$ which is also about 3 times larger in absolute value than
  in the FM cell. Similarly, the interaction between 1 and 4 which also corresponds to 1
  and 2 in the FM cell is even larger at $-0.1538$. All of these values are in mRy. The
  matrix of exchange interactions in this case has the form
\begin{equation}
  J^0=\left(\begin{array}{cccc}a&b&c&d \\b&a&d&c \\c&d&a&b\\d&c&b&a
  \end{array}\right)
\end{equation}
  with $a=-0.107$, $b=-2.464$, $c=-0.231$ and $d=-0.173$. Its largest eigenvalue is
  $a-b-c+d$ and yields a mean-field temperature of 254 K. The Tyablikov approach yields a
  significant reduction to 97 K. The spin arrangement of atoms 1,2,3,4 is
  $\uparrow\downarrow\downarrow\uparrow$, which in fact, the same as the reference state
  we started from, so that now the $T_c$ indeed comes out positive. Also, note that the
  mean-field estimate here is close to the very simple model with an effective $J_{eff}$
  obtained from the AFM-FM total energy difference. That effective interaction represents
  the sum over all individual exchange interactions in the periodic system, in other words
  the $J^0_{12}=\sum_{\bf T}J_{12}^{0{\bf T}}$, excluding the on-site term, which indeed
  is $-2.464$ mRy corresponding to 259 K. So, these different ways of estimating the
  mean-field $T_c$ are roughly consistent with each other. Comparing with the prediction
  starting from the FM reference state, we consistently obtain an AFM ordering along
  alternating (110) planes and all spins in these planes parallel, both in the
  down-pointing and up-pointing square Co pyramids, or the two sites in the primitive
  cell. The mean-field approach gives a substantially larger $T_c$  when starting from the
  AFM reference state, but the final RPA estimates which are expected to give a lower
  bound are not that far from each other 93 K vs. 130 K.  Thus we can safely conclude that
  the N\'eel temperature is approximately 100K.

  The large value of the exchange interaction between Co in line with the O between
  nearest neighbor primitive cells suggests that it is derived from superexchange via the
  O between $d_{x^2-y^2}$ orbitals. The AFM interaction between the two Co within the
  primitive cell on the other hand, is likely to be a direct interaction between $d_{xy}$
  orbitals, which would nearly point to each other except that the Co atoms are in
  slightly different horizontal planes. The $d_{xz},d_{yz}$ orbitals or their
  superposition along a [110] direction could also contribute to this via direct
  antiferromagnetic coupling.

  The magnetic ordering is thus rather interesting with large magnetic moments of 4
  $\mu_B$ interacting differently via the different Co-$d$ orbitals involved. The dominant
  interaction is super-exchange but a smaller direct interaction between atoms in the same
  unit cell also plays a role. One might speculate that spin-fluctuations of this smaller
  interactions combined with doping to create a metallic state, either $p$ or $n$-type
  doping might lead to interesting effect and possibly spin-fluctuations mediated
  superconductivity. 
  

\subsection{AFM band structure and optical properties}
    \begin{figure}
      \includegraphics[width=8cm]{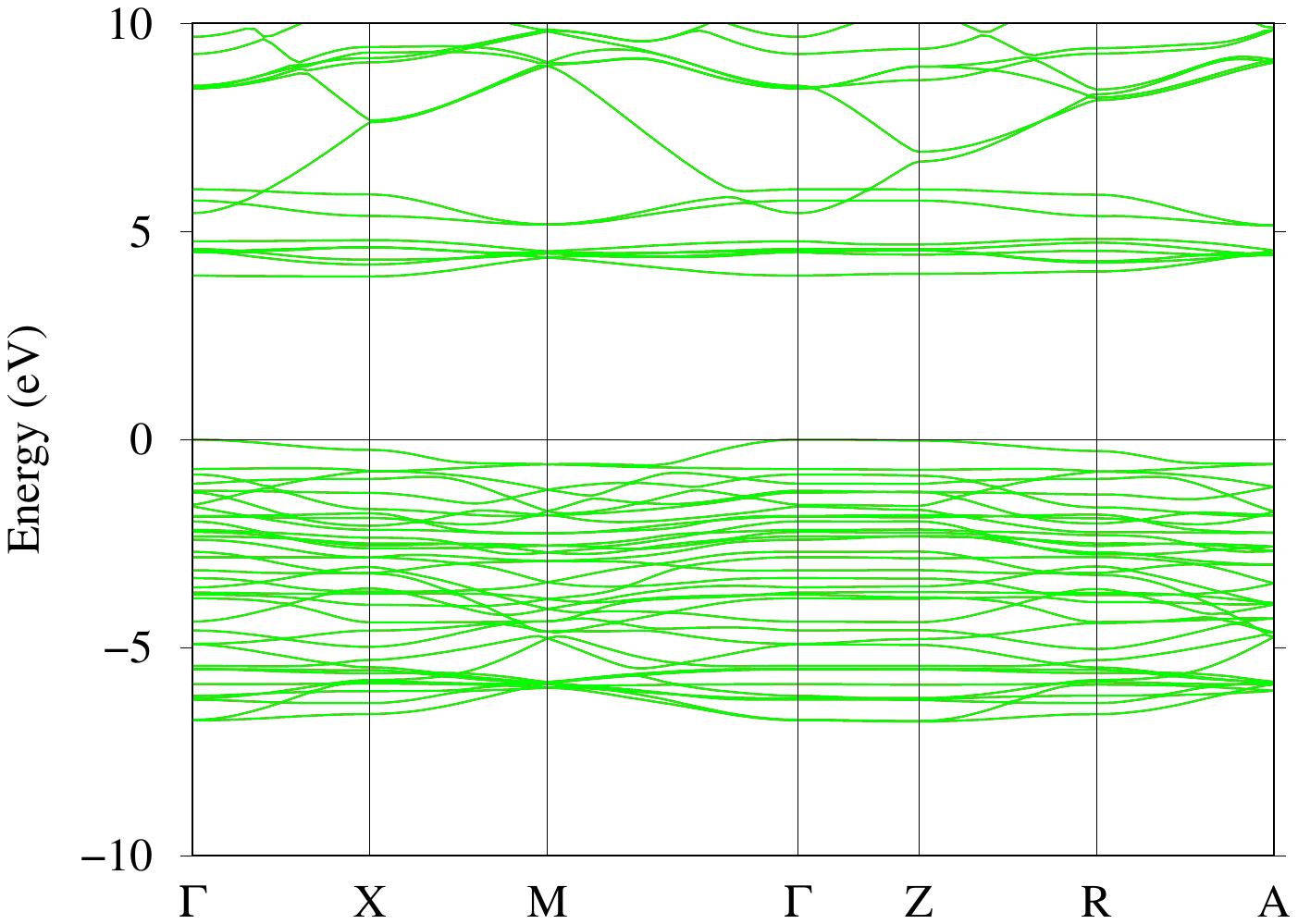}
      \caption{Band structure of AFM KCoO$_2$ int he QS$GW$ approach.\label{figbndsafm}}
    \end{figure}

    \begin{figure}
      \includegraphics[width=6cm]{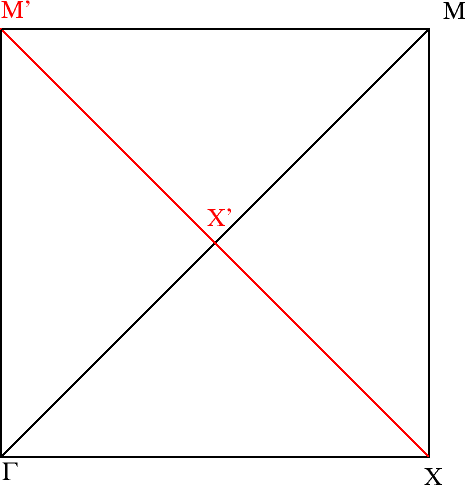}
      \caption{Relation between primitive cell tetragonal Brillouin zone labeled in black
               and doubled cell 45 degree rotated Brillouin zone for AFM case, labeled in
               red and with primes.\label{figbzs}}
    \end{figure}

    \begin{figure}
      \includegraphics[width=8cm]{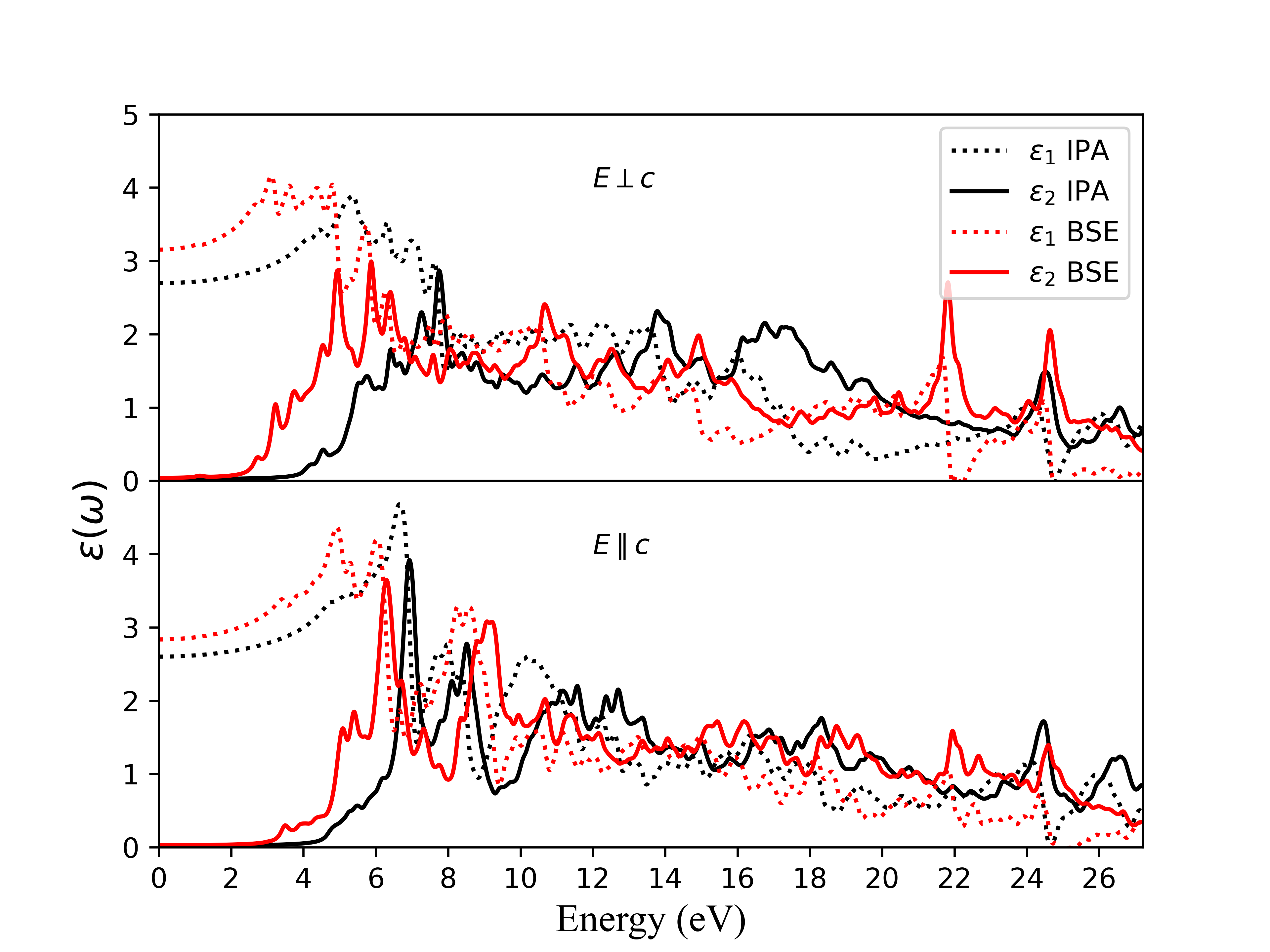}
      \caption{Optical dielectric function for the AFM state.\label{figoptafm}}
    \end{figure}
  The antiferromagnetic band structure is shown in Fig. \ref{figbndsafm}. The relation
  between the FM primitive cell Brillouin zone and the AFM cell Brillouin zone is shown in
  Fig.~\ref{figbzs}. The new $\Gamma X$ direction is half the old $\Gamma M$ direction and
  the bands are essentially folded in two in that direction. The new $M$ point corresponds
  to the old $X$ point. The band structure now has a direct gap at $\Gamma$ of 3.94 eV
  because the spin up VBM state at $M$ becomes folded on the new $\Gamma$ state. Both the
  top valence and bottom conduction bands become almost entirely flat along $\Gamma Z$.

  The optical dielectric function of the AFM state is shown in Fig.~\ref{figoptafm}. It is
  calculated here using $N_v=48$ valence bands and $N_c=40$ conduction bands and a
  $3\times3\times3$ {\bf k}-mesh. It is rather similar to the corresponding FM case shown
  in Fig.~\ref{figeps} although not quite identical. In both cases, we may note a
  substantial redshift between the IPA and BSE $\varepsilon_2(\omega)$ and excitonic
  features below the quasiparticle gap of 3.94 eV. We note that the sharp features for
  ${\bf E}\perp{\bf c}$  at 22 and 24 eV may be artifacts of the truncation of the active
  space in the BSE calculation. For a smaller set, similar features appeared for
  ${\bf E}\parallel{\bf c}$ around 15 eV but these disappear or are weakened when
  more conduction bands were included. The calculation may be deemed to be converged up to
  about 12 eV as in this range they are the same with higher or lower $N_c$.
  
\section{Conclusions}
  The main conclusions from our study are summarized as follows. KCoO$_2$ has large
  magnetic moments of $4\mu_B$ on the Co atoms, corresponding to a $d_{xy}^2$,
  $d_{xz}^1$, $d_{yz}^1$, $d^1_{x^2-y^2}$, $d^1_{3z^2-r^2}$ configuration, arising from
  the square pyramidal coordination with a K ion on the opposing apical site. The magnetic
  moments prefer to order antiferromagnetically along the [110] direction. The exchange
  interactions are dominated by an antiferromagnetic super-exchange coupling between
  $d_{x^2-y^2}$ orbitals $\sigma$-bonding with O-$p$ orbitals between them in a
  180$^\circ$ alignment of order $\sim$8 meV but with smaller direct exchange between the
  two Co per primitive cell, thus aligning all spin on atoms in successive (110) planes.
  The N\'eel temperature was predicted to be about 100 K using the Tyablikov-Callen
  approach and using exchange interactions extracted from the transverse spin
  susceptibility in a rigid spin approximation per sphere and including a converged
  summation of exchange interactions. In the mean-field approximation, a larger critical
  temperature of about 250 K is obtained, which  provides an upper limit. In the
  ferromagnetic state, the band structure exhibits an indirect gap between the conduction
  band minimum at $\Gamma$ of minority spin and a valence band maximum of majority spin
  at $M$, the corner of the Brillouin zone in the $k_z=0$ plane. The optical transitions
  between equal spin were calculated including electron-hole interaction effects and show
  strongly bound excitons. In the AFM case, the band edges of the gap show extremely flat
  regions along the direction perpendicular to the layers and a direct quasiparticle gap
  of 3.94 eV. The combination of large magnetic moments, relatively small exchange
  interactions of different types and flat band edge states indicate that strong
  correlation effects may be present in this system and could lead to interesting effects,
  in particular when doping is considered to modify the antiferromagnetic insulating
  ground state. 
  
\acknowledgements{This work was supported by the U.S. Air Force Office of Scientific
                  Research (AFOSR) under grant no. FA9550-22-1-0201 (Program Manager Ali
                  Sayir), and made use of the High Performance Computing Resource in the
                  Core Facility for Advanced Research Computing at Case Western Reserve
                  University. J.J. acknowledges support under the CCP9 project
                  {\emph{Computational Electronic Structure of Condensed Matter}}
                  (part of the Computational Science Centre for Research Communities
                  (CoSeC)).}



\bibliography{lco,kcoo2,dft,gw,lmto,spinchi,BSE}
\end{document}